\UseRawInputEncoding
\documentclass[submission,copyright,creativecommons]{eptcs}
\providecommand{\event}{Post Proceedings of ThEdu'21} 

\usepackage{xurl}
\usepackage{graphicx}
\usepackage{amsmath}
\usepackage{amsthm}

\usepackage{hyperref}
\usepackage{color}
\usepackage{subcaption}

\usepackage{listings,newtxtt}
\lstdefinelanguage{giac}{
  ndkeywords={>>},
  sensitive=true
}

\lstdefinelanguage{mylog}{
  sensitive=true
}

\title{Symbolic Comparison of Geometric Quantities in GeoGebra}
\author{Zolt\'an Kov\'acs
\institute{The Private University College of Education of the Diocese of Linz\\Linz, Austria}\thanks{The
work was partially supported by the grant 
PID2020-113192GB-I00 from the Spanish MICINN.}
\email{zoltan@geogebra.org}
\and
R\'obert Vajda
\institute{Bolyai Institute, University of Szeged\\ Szeged, Hungary}
\email{vajda@math.u-szeged.hu}
}
\def\titlerunning{Symbolic Comparison in GeoGebra}
\def\authorrunning{Kov\'acs \& Vajda}
\begin{document}
\maketitle

\begin{abstract}
Comparison of geometric quantities usually means obtaining
generally true equalities of different algebraic expressions of a given geometric figure.
Today's technical possibilities already support symbolic proofs of
a conjectured theorem, by exploiting computer algebra capabilities of
some dynamic geometry systems as well. We introduce \textit{GeoGebra}'s new
feature, the \textit{Compare} command, that helps the users in experiments in planar geometry.
We focus on automatically obtaining conjectures and their proofs at the same time,
including not just equalities but inequalities too. Our contribution can already be successfully
used to support teaching geometry classes at secondary level, by getting several well-known and
some previously unpublished result within seconds on a modern personal computer.
\end{abstract}

\bibliographystyle{eptcs}

\tolerance10000

\section{Introduction}

Planar Euclidean geometry always played an important role in history. Euclid's famous book \textit{Elements},
until recent times, was supposedly the second-best selling book of all time, after the Bible.
In the modern era, teaching geometry had different focus in different countries during
different educational and political systems.
But recently the wide-spread of dynamic geometry applications
allowed for today's schools to rediscover the beauty of this topic
(see \cite{Davis95} for a summary), either in frontal teaching,
or in small groups, or even in individual work. Dynamic geometry applications
also influence contemporary research in planar geometry---they can help
finding new theorems (see e.g.~\cite{LRV2011}).

For geometric equalities there are already integrated tools in some dynamic geometry systems (DGS).
Maybe the most well-known application is GeoGebra---it provides an automated reasoning toolset (ART)
that supports symbolic check of equality of
lengths of line segments in a geometric configuration, or, the equality of expressions.
Both features are supported by the \textit{Relation} tool or command,
or the low-level commands, \textit{Prove} and \textit{ProveDetails}, directly \cite{RelTool-ADG2014}.
On one hand, GeoGebra's ART successfully proves hundreds of well-known theorems in
elementary planar geometry\footnote{See
\url{https://prover-test.geogebra.org/job/GeoGebra_Discovery-provertest/69/artifact/fork/geogebra/test/scripts/benchmark/prover/html/all.html}
for a recent benchmark of 291 test cases.}, and, on the other hand, there are examples
of obtaining automated proofs of more difficult, much recent statements%
\footnote{See e.g.~\url{https://matek.hu/zoltan/blog-20201229.php}.},
or of completely new results\footnote{See e.g.~\cite{LNAI11006-isoptics}.}.

In this paper we describe our current work that goes one step forward.
In our work we use partly the same algebro-geometric theory that
plays a crucial role in GeoGebra's toolset, based on the revolutionary work of Wu \cite{Wu78}, Chou \cite{Chou88},
and improved later by Recio and V\'elez \cite{RecioVelez99} with elimination theory.
This is the required algebraic basis for gaining a conjecture on a possibly fixed ratio of two quantities.
Once the elimination method is able to provide the exact ratio, this conjecture is in fact a mathematically proven proposition.

On the other hand, we use partly general purpose real quantifier elimination (RQE)
methods to find the best possible geometric constants in the related inequalities
if two expressions do not have a fixed ratio. This approach supports studying
the class of general non-degenerate triangles, or a more specific class of triangles such as isosceles triangles.

In this study we do not go into the further detail on technical difficulties, but refer
to two recent papers: \cite{mcs2020} explains
the mathematical background on obtaining ratios in regular polygons,
while \cite{scsc-2020} summarizes the RQE related issues,
and points to a large set of benchmarks based on our tool. Instead, our current paper is rather focusing
on the practical use: how our work can be fruitful for the student, the teacher and the researcher.
We present an experimental version of GeoGebra, namely,
\textit{GeoGebra Discovery}, which tries to solve the problem with elimination first,
and if this step is unsuccessful, then the given RQE problem setting will be outsourced
to an external tool \textit{realgeom}. We highlight that realgeom can work together with Wolfram's \textit{Mathematica}
\cite{Mathematica}
(in this case the end user is expected to have a Mathematica subscription, or to own
a \textit{Raspberry Pi} system that offers free access to Wolfram's tool, or download a
free developer command-line version of Mathematica called \textit{WolframScript}),
or the free system \textit{Tarski}
\cite{ValeEnriquez-Brown}
(which outsources some computations to the \textit{QEPCAD B} system \cite{Brown03}).

Our experimental system is already capable
of solving a large set of open questions in planar Euclidean geometry. We believe that
our contribution to automated reasoning can be fruitful for several levels of mathematics
education. What is more, researchers can double check new conjectures and may obtain even some
non-trivial or yet unknown results.

We emphasize that the mathematical methods we use in this paper are mostly well-known,
however, a full implementation of the various techniques in a single graphical application,
being freely available for millions of potential users, is completely new.

\section{Comparison of expressions at school}

School curriculum usually includes several relationships between expressions in a planar construction.
Every student has to learn the Pythagorean theorem which is an equality of two expressions,
namely $a^2+b^2$ and $c^2$ where $a$, $b$ and $c$ correspond to the lengths of sides in a right triangle.
In general, however, a simple question is not discussed: what happens if we omit the assumption
that the triangle has a right angle? By using our tool it can be quickly shown mechanically that in general the
inequality $\displaystyle{a^2+b^2>\frac{c^2}2}$
holds.

In practical life everybody solves inequalities. Given the task that somebody from point $A$ has to reach
point $B$ as far as possible, no external point $C\not\in AB$ will be chosen to visit that
point first and just then $B$---unless the terrain is not planar.
Naturally, the shortest path is ``as the crow flies'', or, in mathematical means:
the triangle inequality $|AB|\leq|AC|+|BC|$ always holds.

Mathematics curriculum seems to have less to do with inequalities. This is especially true for
the subtopic of geometry.
We rather prefer relationships that express equality. To name a few, the
\textit{intercept theorem} or the \textit{geometric mean theorem} can
be mentioned (the latter is a special case of the \textit{chord theorem}), or, at a higher level,
Ptolemy's theorem,
the Ceva-Menelaus formulas or Heron's formula may appear in certain school types.
For inequalities we can think of the well-known correspondence between the angles and sides
in a triangle (namely, $a\leq b\Leftrightarrow\alpha\leq\beta$), or, as a kind of generalization of
Thales' circle theorem, the inscribed angle theorem gives a suggestion how to compare
angles lying on different arcs that share the same chord.
From advanced topics we can mention \textit{Euler's inequality} (1765, but found earlier by Chapple in 1746,
\cite{Chapple1746})
between the radii of the circumcircle
and incircle of a triangle, or the \textit{Erd\H{o}s-Mordell inequality} (1937, \cite{Mordell1937}).

These examples are well-known theorems. In fact, comparison of given quantities is a more general task,
and can be a very rich field of new experiments, exercises, problems or even new theorems.
On the other hand, equality-based problems are much more usual in the literature.
For example, by having a deeper look on the problem lists of International Mathematical Olympiad
for the last 20 years, very few geometrical proofs of inequalities were given, even if one-third of the problem
settings are traditionally related to planar Euclidean geometry.
(See \url{https://www.imo-official.org/problems.aspx} for a complete list. In particular,
problems 2001/1, 2002/6, 2006/1, 2013/2 and 2020/6 are related to geometrical inequalities.)

We need to admit that a certain type of inequalities has already been studied quite exhaustively
during the last decades. Here the main reference is, without any doubt, \cite{Bottema69}.
In a part of that work several inequalities of homogeneous type are considered, like
\begin{equation}
3\cdot(ab+bc+ca)\leq(a+b+c)^2<4\cdot(ab+bc+ca)
\label{bottema11}
\tag{Bottema 1.1}
\end{equation}
where $a$, $b$ and $c$ denote the lengths of the sides of an arbitrary
non-degenerate triangle in the plane. In general we can reformulate the
above result to an open question, namely: Are there
sharp constants $m,M$ such that
\begin{equation}
m\cdot(ab+bc+ca)\underset{(=)}{<}(a+b+c)^2\underset{(=)}{<}M\cdot(ab+bc+ca)?
\label{bottema11-q}
\end{equation}
The answer is, according to \cite{Bottema69}, yes, namely $m=3$ and $M=4$.

To \textit{mechanically} give an exact answer we already pointed to the RQE approach.
But we will also focus on the ``$m=M$'' case in our algorithm---that is, we assume that
for the input expressions $w_1$ and $w_2$ there exists a positive number $\mu$ such that
\begin{equation}
\mu\cdot w_1=w_2
\end{equation} always holds, for (almost) all possible positions of the free points in the input figure.
(In our example above $w_1=ab+bc+ca$ and $w_2=(a+b+c)^2$.)
Of course, this is often not the case, for example, in (\ref{bottema11-q}) not, since $\mu\in[3,4)$,
that is, $m=3$ and $M=4$.

Here we highlight that
the case ``$m=M$'' can usually be found by utilizing elimination theory as well.
This remark can be extremely important to reduce the computational complexity of the mechanical solution
in many cases. In the next section, after discussing an example when $m\neq M$ (Fig.~\ref{figure:bottema11-gd})
we will show a non-trivial example for the case $m=M$ (Fig.~\ref{figure:relation-pentagon1},
\ref{figure:relation-pentagon2} and \ref{figure:relation-pentagon3}).

\section{The Compare command in GeoGebra Discovery}

\begin{figure}
\begin{center}
\includegraphics[width=\textwidth]{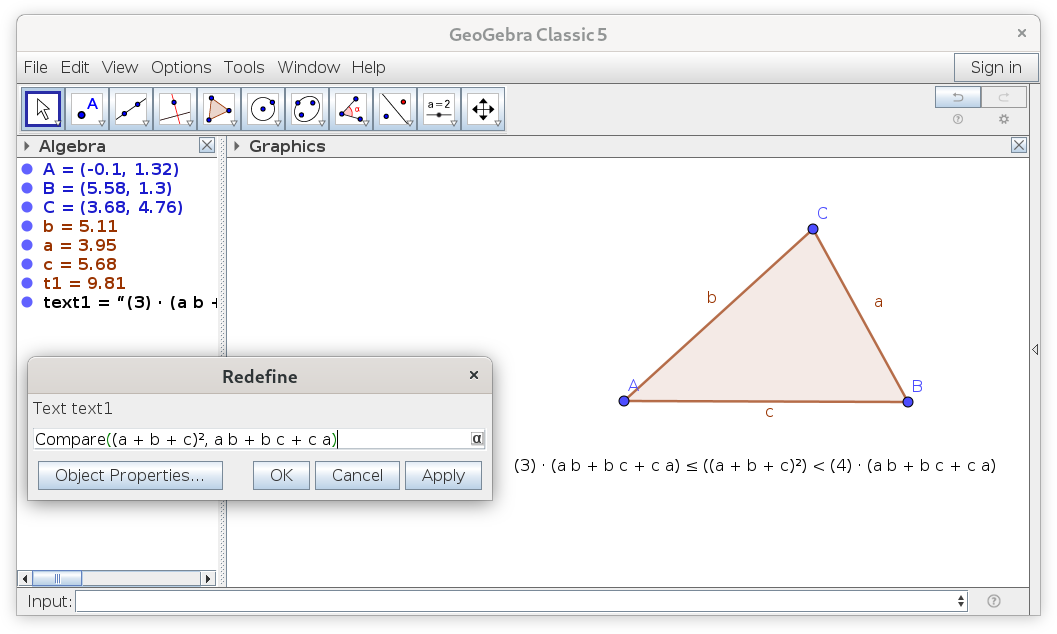}
\caption{A mechanical solution by GeoGebra Discovery for the problem Bottema 1.1}\label{figure:bottema11-gd} 
\end{center}
\end{figure}

As a first example we show how the question (\ref{bottema11-q}) can be asked in a recent version
\texttt{2021Sep03} of GeoGebra Discovery\footnote{\url{https://github.com/kovzol/geogebra-discovery}}.
The user constructs a triangle $ABC$. Implicitly the sides $a$, $b$ and $c$ will also be added by the program.
Now the command \texttt{Compare($(a+b+c)^2$,$ab+bc+ca$)} is to be typed and the program
shows the algebraic result in the bottom of the \textit{Graphics View} as seen in Fig.~\ref{figure:bottema11-gd}.

In fact, GeoGebra Discovery is programmed to realize that the expressions $(a+b+c)^2$ and $ab+bc+ca$ both are
homogeneous quadratic formulas. 

\begin{figure}
\begin{center}
\includegraphics[width=\textwidth]{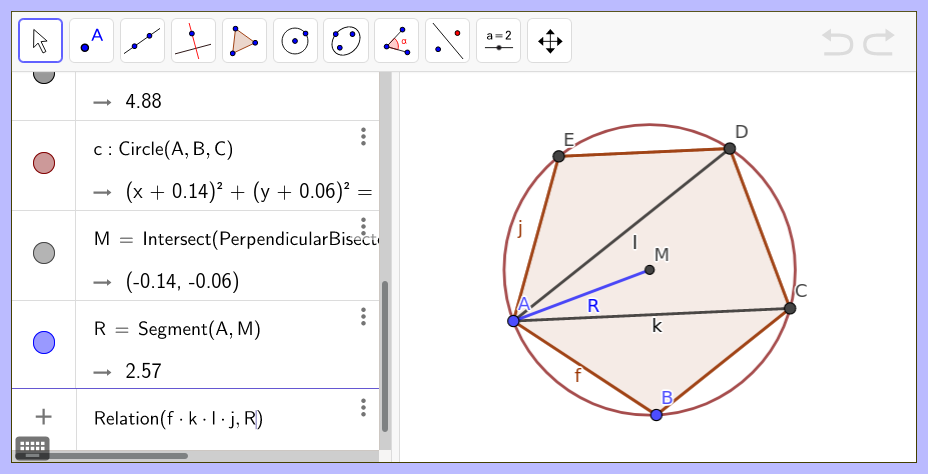}
\caption{Using the Relation command to obtain a relationship in a regular pentagon}\label{figure:relation-pentagon1} 
\end{center}
\end{figure}

\begin{figure}
\centering
\begin{subfigure}[b]{0.4\textwidth}
\centering
\includegraphics[width=\textwidth]{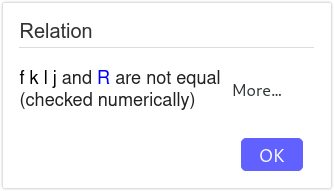}
\caption{Numerical check}\label{figure:relation-pentagon2} 
\end{subfigure}
\hfill
\centering
\begin{subfigure}[b]{0.4\textwidth}
\includegraphics[width=\textwidth]{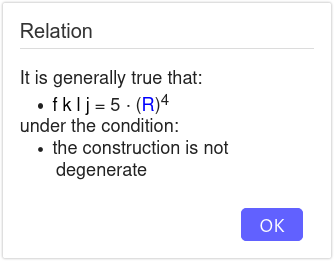}
\caption{Symbolic check}\label{figure:relation-pentagon3} 
\end{subfigure}
\caption{Results on comparing $f\cdot k\cdot l\cdot j$ and $R$}
\label{checks}
\end{figure}

We illustrate the same idea on an equality in a regular pentagon. In Fig.~\ref{figure:relation-pentagon1}
a regular 5-gon $ABCDE$ is shown with the midpoint $M$ of its circumcircle. We introduce the
quantities $f=AB$, $k=AC$, $l=AD$, $j=AE$, $R=AM$. After some experiments one can find that
there is an interesting relationship among the product $f\cdot k\cdot l\cdot j$ and $R$. In this example,
however, we use the command \texttt{Relation($f\cdot k\cdot l\cdot j$,$R$)} to do a numerical check
first (Fig.~\ref{figure:relation-pentagon2}), and also to have a nicely looking symbolic result when clicking on the
button ``More$\ldots$'' (Fig.~\ref{figure:relation-pentagon3}).

\begin{figure}
\begin{center}
\includegraphics[width=0.6\textwidth]{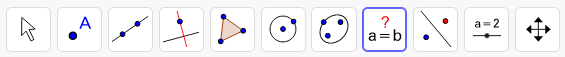}
\caption{The Relation tool in the toolbar in GeoGebra 6}\label{figure:toolbar} 
\end{center}
\end{figure}

\begin{figure}
\begin{center}
\includegraphics[width=0.4\textwidth]{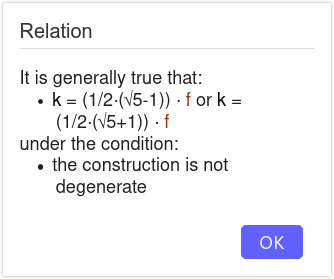}
\caption{Symbolic relationship between the side and a diagonal of a regular pentagon}\label{figure:relation-pentagon4} 
\end{center}
\end{figure}

The Relation command in GeoGebra is a well-known way to compare two objects, from its first versions,
but the comparison used to be performed only numerically. If one wants to compare just two
objects (that is, not expressions, but, for example, segments), then a point-and-click comparison
is also possible. This is demonstrated in Fig.~\ref{figure:toolbar}: after selecting the Relation tool,
two segments, e.g.~$f$ and $k$, we learn that their ratio is
$\frac12\cdot(\sqrt5-1)$ or $\frac12\cdot(\sqrt5+1)$
(see Fig.~\ref{figure:relation-pentagon4}). Confusingly, it remains an unanswered question here whether
the first or the second one is the correct result.

To explain this issue, we need to distinguish between two main GeoGebra versions: 5 and 6. Version 5 has support to
handle inequalities, but version 6 not at the moment. In version 6
the underlying computations
in the algebro-geometric setup  cannot restrict the geometric figure to a single regular pentagon.
Instead of a single regular polygon, all possible star-regular polygons are observed at the same time.
In the case of a regular pentagon this means that the computation for a regular pentagram will
also be computed---two results will be given concurrently.
In general this means that the final result always includes a solution according each star-regular variant.
(Fortunately, for a square or a regular hexagon there are no star-regular variants, but for a
regular heptagon there are \textit{two} heptagrams---so three cases are computed at the same time.)

\begin{figure}
\begin{center}
\includegraphics[width=0.4\textwidth]{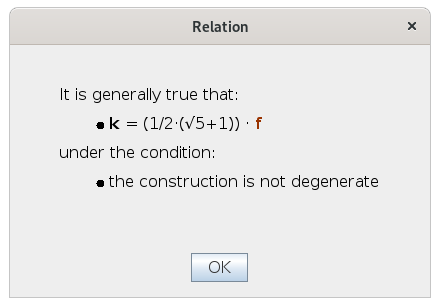}
\caption{Unambiguous result on the ratio between the side and a diagonal of a regular pentagon (supported only in GeoGebra 5)}\label{reg5d} 
\end{center}
\end{figure}

In version 5 we are already capable of issuing a restriction that point $C$ takes place in the half-plane whose
boundary line includes $B$ and is perpendicular to the segment $AB$.
In this case we can exclude the pentagram variant (Fig.~\ref{reg5d}). With such considerations
we can extend the equation based approach to a much broader set of applications. Among others,
it is possible to attach a point on a segment or inside a triangle, or to explicitly define the
center of the incircle of a triangle.

In the beginning of this section we mentioned the Compare command in GeoGebra Discovery,
and finally we arrived to a higher level command/tool: Relation. In fact, Relation uses the
internal routine of the Compare command, so, at the end of the day, the same computations
are performed, independently of the launched command.

\begin{figure}
\begin{center}
\includegraphics[scale=0.7]{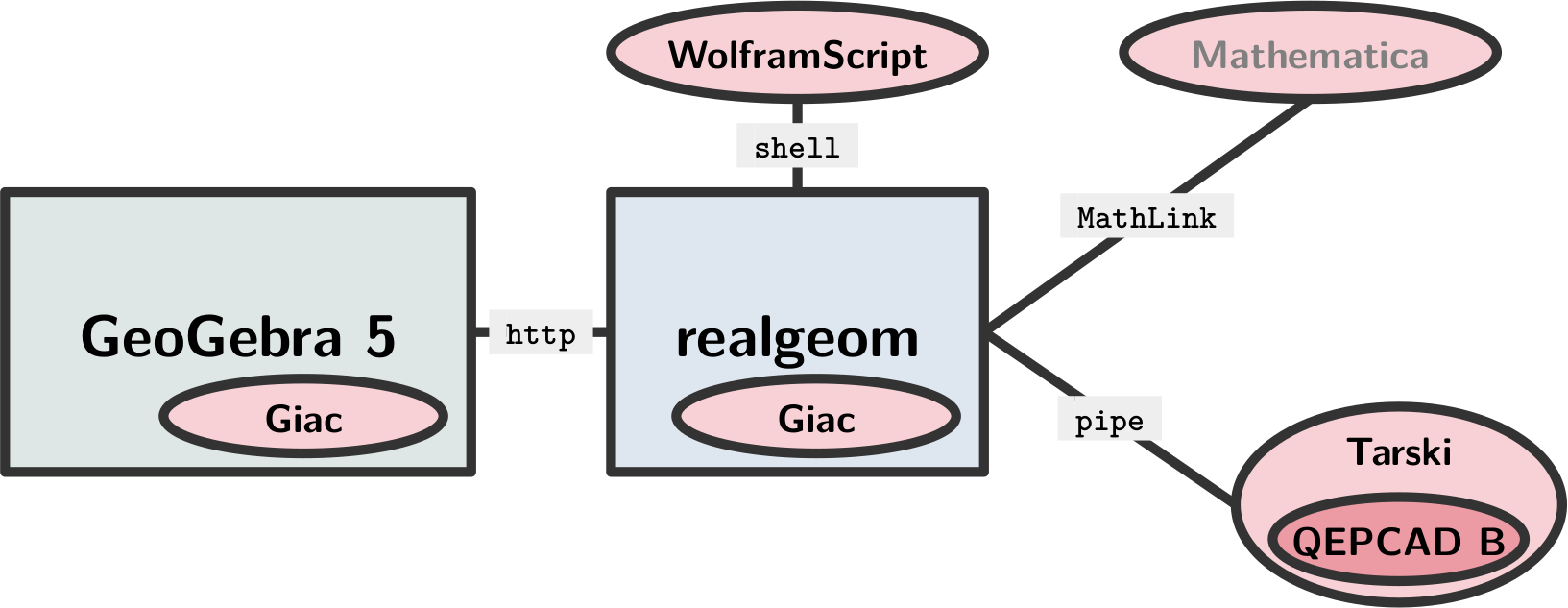}
\caption{Computation hierarchy for GeoGebra 5}\label{gg-rg3} 
\end{center}
\end{figure}

Finally we give some more description on the system layout of the above mentioned two GeoGebra versions, 5 and 6.
At the moment only GeoGebra 5 (a native Java application)
can give any result on inequalities, because all RQE computations are outsourced to Mathematica or Tarski.
Fig.~\ref{gg-rg3} gives an overview of the underlying technical hierarchy. 
GeoGebra internally uses the Giac computer algebra system to perform a large set of symbolic computations
\cite{GiacGG-RICAM2013}.

\begin{figure}
\begin{center}
\includegraphics[scale=0.6]{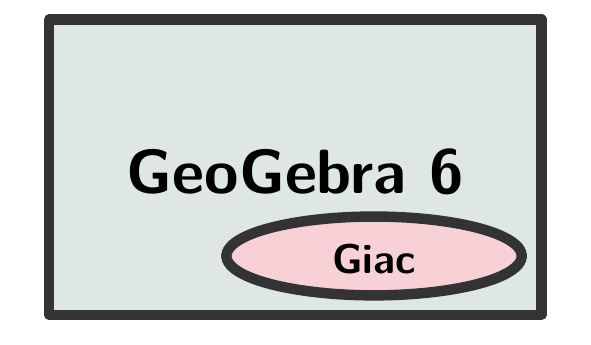}
\caption{Computation hierarchy for GeoGebra 6}\label{gg-rg3b} 
\end{center}
\end{figure}

By contrast,  GeoGebra 6 (a web browser based variant of GeoGebra) is capable only of obtaining
\textit{equalities}, because no connection has been made yet to any systems that provide RQE computations.
See Fig.~\ref{gg-rg3b} for the simple schema being available for the moment.

\section{Mathematical background}

We point the reader to the first experiments on this topic to \cite{SturmWeispfenning96} (on proving
inequalities) and \cite{RecioVelez99} (on obtaining equalities). By using a non-trivial
but still simple example we explain the main idea how our algorithms work.

\begin{figure}
\begin{center}
\includegraphics[scale=0.4]{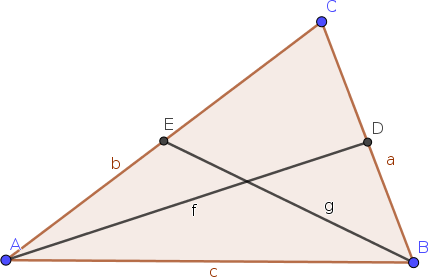}
\caption{Problem setting}\label{ex4-1} 
\end{center}
\end{figure}

\begin{figure}
\begin{center}
\includegraphics[scale=0.4]{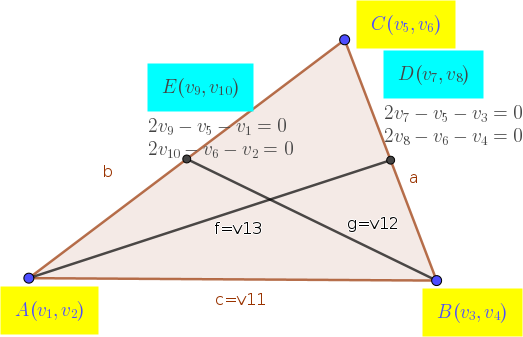}
\caption{Algebraization of the problem setting}\label{ex4-2} 
\end{center}
\end{figure}

The problem setting, as visible in Fig.~\ref{ex4-1}, is to compare $f+g$ and $c$ in an arbitrary
triangle $ABC$---here $f$ and $g$ are two medians of the triangle, and $c$ is the side of the triangle
exactly as drawn in the figure.
Now Fig.~\ref{ex4-2} shows how this problem is translated into an algebraic setup:
each of the free points $A$, $B$ and $C$ are expressed by two coordinates ($v_1,v_2,v_3,v_4,v_5,v_6$), dependent points
$D$ and $E$ are given with another two times two coordinates ($v_7,v_8,v_9,v_{10}$), and another three variables
describe the length of $c$, $g$ and $f$ ($v_{11},v_{12},v_{13}$). All of these mappings are done
automatically by GeoGebra by following a mechanical algorithm, and also some equations are
set up to describe the relationships between these variables---in our case four equations will be created.

Four or five more technical variables
are introduced:
\begin{itemize}
\item $v_{14}$ is used to prescribe a condition on the non-degeneracy of $ABC$, namely,
$$v_{14}\cdot\left|\begin{matrix}v_1&v_2&1\\v_3&v_4&1\\v_5&v_6&1\end{matrix}\right|-1=0,
$$ by using the Rabinowitsch trick \cite{Rabinowitsch1929}. Here the geometrical
meaning is an inequation, even if we use an \textit{equation} for its algebraic translation.
This ensures that only algebraic equations are set up in the first round and no
algebraic inequality will be present. To use effective elimination via Gr\"obner bases this
step cannot be avoided.
\item $w_1$ and eventually $w_2$ are used to describe the two expressions with a single variable, respectively.
\item $m$ plays the role of $m$, $M$ and $\mu$ as explained above.
\item $n$ is used to express that one of the two expressions differs from $0$. Here, again,
the Rabinowitsch trick is used.
\end{itemize}

After these settings GeoGebra uses elimination first if a suitable $\mu$ exists. To achieve this,
the following Giac code is executed:
\begin{lstlisting}[language=giac]
[assume(m>0),solve(eliminate(subst([2*v7-v5-v3,2*v8-v6-v4,2*v9-v5-v1,2*v10-v6-v2,-v12^2+v10^2+v9^2-2*v10*v4+v4^2-2*v9*v3+v3^2,-v11^2+v4^2+v3^2-2*v4*v2+v2^2-2*v3*v1+v1^2,-v13^2+v8^2+v7^2-2*v8*v2+v2^2-2*v7*v1+v1^2,-1-v14*v5*v4+v14*v6*v3+v14*v5*v2-v14*v3*v2-v14*v6*v1+v14*v4*v1,-w1+(v13+v12)^1,v11*m-(w1),(v11)*n-1],[v1=0,v2=0,v3=1,v4=0]),[v1,v2,v3,v4,v5,v6,v7,v8,v9,v10,v11,v12,v13,v14,w1,n])[0],m)][1]
\end{lstlisting}
We note that the first free two points are positioned to $(0,0)$ and $(1,0)$, respectively. This is
allowed without loss of generality---with the assumption that the algorithm always verifies if the expressions
are homogeneous and of the same degree. (See also \cite{Davenport2017}.)

If Giac gave a non-erroneous output here, then we managed to find the sought $\mu$
(or, eventually a set of possible values for $\mu$). Otherwise
there is a second round needed---like in our case in this example. GeoGebra now outsources
the problem to a realgeom server, by composing an http request, something like
\nolinkurl{http://192.168.10.20:8765/euclideansolver?lhs=w1&rhs=v11&polys=2*v7-v5-v3,2*v8-v6-v4,2*v9-v5-v1,2*v10-v6-v2,-v12%5E2%2Bv10%5E2%2Bv9%5E2-2*v10*v4%2Bv4%5E2-2*v9*v3%2Bv3%5E2,-v11%5E2%2Bv4%5E2%2Bv3%5E2-2*v4*v2%2Bv2%5E2-2*v3*v1%2Bv1%5E2,-v13%5E2%2Bv8%5E2%2Bv7%5E2-2*v8*v2%2Bv2%5E2-2*v7*v1%2Bv1%5E2,-1-v14*v5*v4%2Bv14*v6*v3%2Bv14*v5*v2-v14*v3*v2-v14*v6*v1%2Bv14*v4*v1,-w1%2B(v13%2Bv12)%5E1&vars=v1,v2,v3,v4,v5,v6,v7,v8,v9,v10,v11,v12,v13,v14,w1&posvariables=v12,v13,v11&mode=explore&timelimit=5}.
This triggers realgeom's internal computations that finally launch a request towards Mathematica (or Tarski)
like this:
\begin{lstlisting}[language=mylog]
2021-02-01 14:16:42.254 LOG: ineqs=(m>0)
2021-02-01 14:16:42.254 LOG: before substitution, polys=2*v7-v5-v3,2*v8-v6-v4,2*v9-v5-v1,2*v10-v6-v2,-v12^2+v10^2+v9^2-2*v10*v4+v4^2-2*v9*v3+v3^2,-v11^2+v4^2+v3^2-2*v4*v2+v2^2-2*v3*v1+v1^2,-v13^2+v8^2+v7^2-2*v8*v2+v2^2-2*v7*v1+v1^2,-1-v14*v5*v4+v14*v6*v3+v14*v5*v2-v14*v3*v2-v14*v6*v1+v14*v4*v1,-w1+(v13+v12)^1
2021-02-01 14:16:42.255 LOG: before delinearization, polys=2*v7-v5-1,2*v8-v6,2*v9-v5,2*v10-v6,-v12^2+v10^2+v9^2-2*v9+1,-v11^2+1,-v13^2+v8^2+v7^2,-1+v14*v6,v13+v12-w1,(w1)-m*(v11)
2021-02-01 14:16:42.255 LOG: delinearization code=...
Input: 10 eqs in 12 vars
Considering positive roots of -v11^2+1=0 in variable v11
{v11=-1,v11=1}
Positive root is 1
New set: {-v5+2*v7-1,-v6+2*v8,-v5+2*v9,2*v10-v6,v10^2-v12^2+v9^2-2*v9+1,-v13^2+v7^2+v8^2,v14*v6-1,v12+v13-w1,-m+w1}
Keeping 1-v11
Removing -v5+2*v7-1, substituting v5 by 2*v7-1
New set: {-v6+2*v8,-2*v7+2*v9+1,2*v10-v6,v10^2-v12^2+v9^2-2*v9+1,-v13^2+v7^2+v8^2,v14*v6-1,v12+v13-w1,-m+w1}
Removing -v6+2*v8, substituting v6 by 2*v8
New set: {-2*v7+2*v9+1,2*v10-2*v8,v10^2-v12^2+v9^2-2*v9+1,-v13^2+v7^2+v8^2,2*v14*v8-1,v12+v13-w1,-m+w1}
Removing -2*v7+2*v9+1, substituting v7 by -1/2*(-2*v9-1)
New set: {2*v10-2*v8,v10^2-v12^2+v9^2-2*v9+1,-v13^2+v8^2+v9^2+v9+1/4,2*v14*v8-1,v12+v13-w1,-m+w1}
Removing 2*v10-2*v8, substituting v10 by v8
New set: {-v12^2+v8^2+v9^2-2*v9+1,-v13^2+v8^2+v9^2+v9+1/4,2*v14*v8-1,v12+v13-w1,-m+w1}
Set after delinearization: {-v12^2+v8^2+v9^2-2*v9+1,-v13^2+v8^2+v9^2+v9+1/4,2*v14*v8-1,v12+v13-w1,-m+w1,1-v11}
Delinearization output: 6 eqs in 8 vars
2021-02-01 14:16:42.262 LOG: after delinearization, polys={-v12^2+v8^2+v9^2-2*v9+1,-v13^2+v8^2+v9^2+v9+1/4,2*v14*v8-1,v12+v13-w1,-m+w1,1-v11}
2021-02-01 14:16:42.262 LOG: before removing unnecessary variables, vars=v1,v2,v3,v4,v5,v6,v7,v8,v9,v10,v11,v12,v13,v14,w1
2021-02-01 14:16:42.263 LOG: after removing unnecessary variables, vars=[v12,v8,v9,v13,v14,w1,m,v11]
2021-02-01 14:16:42.263 LOG: after removing m, vars=v12,v8,v9,v13,v14,w1,v11
2021-02-01 14:16:42.263 LOG: code=ToRadicals[Reduce[Resolve[Exists[{v12,v8,v9,v13,v14,w1,v11},(m>0) \[And] (v12>0) \[And] (v13>0) \[And] (v11>0) \[And] (-v12^2+v8^2+v9^2-2*v9+1==0) \[And] (-v13^2+v8^2+v9^2+v9+1/4==0) \[And] (2*v14*v8-1==0) \[And] (v12+v13-w1==0) \[And] (-m+w1==0) \[And] (1-v11==0)],Reals],Reals],Cubics->False]
2021-02-01 14:16:42.407 m > 3/2
\end{lstlisting}
The final result will be actually returned by realgeom to GeoGebra via the http string response \texttt{m > 3/2}.
This is finally communicated by GeoGebra as output $(f + g) > (3/2) \cdot c$.

The whole process is performed in 173 ms on a typical workstation---here we highlight that the longest
part of the computation was spent in Mathematica (143 ms) in the RQE part. In fact, the bottleneck of our algorithm
is indeed the RQE computation, and it can be extremely slow even for some simple inputs. Thus,
before starting the Mathematica computation, some steps are performed to delinearize
the formulas, that is, to decrease the number of variables as much as possible---here
we managed to avoid 1/3 of the variables (8 instead of 12).\footnote{Here we mention that some of the remaining
time is spent with communication between GeoGebra and realgeom via the http protocol,
and between realgeom and the computational backends, but they are not significant since they take just a
couple of milliseconds.}

We do not go into further details regarding the above algorithm.
It can be fully observed in the freely available source code of GeoGebra \cite{geogebra-discovery} and realgeom
\cite{realgeom},
in particular in files \texttt{AlgoCompare.java} and \texttt{Compute.java}, respectively.

\section{Discussion}

At the beginning of this section we emphasize that
automated collection of minimal definition of the geometric figure, and its full translation
to an algebraic counterpart, is a very delicate task. It is very easy to exhaust resources
if the number of variables or the fashion of equations (or inequalities) is unexpected.
This results in a serious limit of using our tool, most notably for inequalities. Although we
already put much effort in optimizing the input for the RQE part, still much work is needed
to have a fast enough tool for a much larger set of supported experiments.

\begin{figure}
\begin{center}
\includegraphics[scale=0.4]{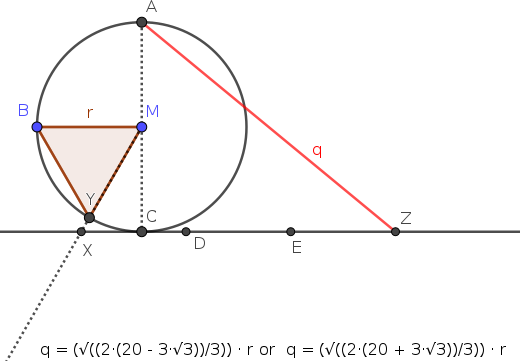}
\caption{GeoGebra finds the symbolic form of Kocha\'nski's approximation for $\pi\approx\sqrt{40/3-2\sqrt3}$,
see also 
\texttt{\href{https://de.wikipedia.org/wiki/N\%C3\%A4herungskonstruktion\_von\_Kocha\%C5\%84ski}{https://de.wikipedia.org/wiki/N\"aherungskonstruktion\_von\_Kocha\'nski}}
}
\label{kochanski} 
\end{center}
\end{figure}

On the other hand, the simpler part (that is, when a constant $\mu$ can be obtained)
can already be used to find complicated relationships as well. A historical example,
an approximation for the number $\pi$, found by Adam Kocha\'nski in 1685 \cite{Kochanski}, can be quickly
obtained by the elimination approach (in 16 ms inside GeoGebra via Giac), however, here also the RQE
method is very powerful (36 ms via realgeom/Mathematica through http) as seen in Fig.~\ref{kochanski}.

That is, a clear benefit of finding $\mu$ is not just to get the required constant,
but also to obtain it in a symbolic form. This can be especially fruitful in education
at geometry
classes where exact computation promotes precise calculations and observing
mathematics as a theoretical science.
It could help understanding the difference between considering mathematics
as just a tool for engineers or
something more sophisticated: a theory that points to \textit{perfection}
by describing numbers that have infinitely many digits---just by finite algorithms.
For example, concerning Kocha\'nski's result the number $3.1415333\ldots$ is just an approximation,
but the ``algorithm'' $\sqrt{40/3-2\sqrt3}$ is the accurate solution.

Even if the RQE part suffers from computational difficulties, there are already
a plenty of possible use cases for the researcher's interest. First of all, several
tests from \cite{Bottema69} can be successfully run for general triangles.
For example, by using the Mathematica backend, Bottema's
1.1--9, 1.12, 1.15, 1.17, 1.19 and 1.23 (14 inequalities) run below 2 seconds.
By contrast, the timing of test 1.14, namely
\begin{align*}
(a + b + c) a^2 b^2 c^2 \leq a b c(a^2 b^2 + b^2 c^2 + c^2 a^2),
\end{align*}
takes more than 5 seconds: even if there are only 148 ms spent in the computation of the RQE step,
the time spent previously for checking if an equational relationship holds via elimination
(which is obviously not the case here), took significantly longer.
In fact, GeoGebra aborts computations
after a 5 seconds of default timelimit---recall that the biggest part of possible users are
students and teachers who do not want to spend too much time (up to hours or days)
with an experiment to get a reasonable answer. For expert users, however, this behavior can be fine-tuned
by using the command line option \texttt{--realgeomws=timeout:$T$} where $T$ is
the maximal amount of allowed time (in seconds). The Tarski backend can successfully
handle 9 inequalities of the above, in a reasonable time too.

On the other hand, much more promising results can be obtained if the general
triangle is changed to some special case. Very fast results can be found
for right triangles---here we tried inequalities 1.1, 1.19--20, 4.2, 5.3, 6.1 and 8.1,
however, in the last three tests it was crucial how the geometric construction
was exactly created. The same set of tests produced very quick results for
isosceles triangles (all below 1 second). The fastest results were obtained
for the special case of an isosceles, right triangle---all of the above tests
performed below 200 ms for this last case.

\begin{figure}
\begin{center}
\includegraphics[scale=1.6]{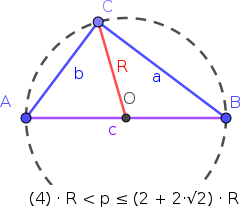}
\caption{Test Bottema 5.3 in a right triangle}\label{b53} 
\end{center}
\end{figure}

We highlight that in our test database (as well as in \cite{Bottema69})
some input expressions contain other, from the geometric point of view more interesting variables than just the
length of the sides. 
For example, the test Bottema 5.3 compares the perimeter $p$ and the circumradius $R$ 
in an arbitrary right triangle (see Fig.~\ref{b53}).

The option to use any variable that refers to a constructed segment of the figure
extends the possibilities to compare a large set of expressions in GeoGebra---practically
by using just a couple of mouse clicks and nothing else. This could be very useful
in education where precise algebraization of the geometric construction
or programming in a computer algebra system
cannot be expected---neither for the student nor the teacher. Also, for a researcher
it can be helpful to use an all-in-one application, at least for the first experiments.

All benchmark data mentioned in our paper above can be downloaded from the webpage
\url{https://prover-test.geogebra.org/job/GeoGebra_Discovery-comparetest/117/artifact/fork/geogebra/test/scripts/benchmark/compare/html/all.html}
(as of 16 September 2021).

\section{Conclusion}

We presented some examples of using our recent improvements on GeoGebra,
namely on comparing geometric expressions symbolically. We think that
our system is already mature enough to be used by a larger set of users.

Further refinements seem to be however beneficiary to enlarge the possible
use cases to a wider set of problems. Here we remark that the general
case of a triangle currently demands too complex computations for many simple inputs.
Moreover, several classic theorems like the Erd\H{o}s-Mordell inequality are still open problems in the general case.
Here we note that the Tarski backend is able to prove Euler's inequality within 1 minute,
but this is perhaps still not adequate in a classroom situation.

\section{Acknowledgments}

We are thankful to Tom\'as Recio and Christopher W.~Brown for their continuous support. We acknowledge
the technical help of the Research Institute for Symbolic Computation (RISC) at JKU Linz, Austria,
by allowing access for us to their Mathematica installations. Finally,
we are grateful to Daniel Carvalho (Wolfram Research) who helped us to improve
the speed of realgeom significantly.

\bibliography{kovzol,external}

\end{document}